\documentstyle[12pt,twoside,fleqn,espcrc1,epsf]{article}

\newcommand{\AmS}{{\protect\the\textfont2
A\kern-.1667em\lower.5ex\hbox{M}\kern-.125emS}}
\def\simlt{\rlap{\lower 3.5 pt\hbox{$\mathchar \sim$}}\raise 1pt \hbox {$<$}}
\def\simgt{\rlap{\lower 3.5 pt\hbox{$\mathchar \sim$}}\raise 1pt \hbox {$>$}}

\title{%
\vspace{-2.8cm}
\begin{flushleft}
       {\normalsize UTCCP-P-63,\ UTHEP-401}   \\[-0.2cm]
       {\normalsize March 1999}   \\
\end{flushleft}
Hadron spectroscopy from lattice QCD}

\author{T. Yoshi\'e\address{
Institute of Physics and Center for Computational Physics,\\
University of Tsukuba, Tsukuba, Ibaraki 305-8577, Japan}
\thanks{
talk presented at KEK-Tanashi International Symposium on
``PHYSICS OF HADRONS AND NUCLEI'' held
at Tokyo, Japan, on December 14-17, 1998.}
}

\begin{document}
\maketitle

\begin{abstract}
I present recent developments in the lattice QCD calculations 
of the light hadron spectrum.
Emphasis is placed on the limitation of the quenched approximation in 
reproducing the observed spectrum and indications that 
the discrepancy is reduced by introducing 
two flavors of light dynamical quarks.
\end{abstract}

\section{INTRODUCTION}
QCD is believed to be the fundamental theory of strong interactions.
This recognition first came from perturbation theory, 
which enables us to describe hadronic processes with large momentum transfer.
However, the conventional perturbative approach fails at 
low energies $p \simlt$ 1 GeV, 
where the strong coupling constant is of order unity. 

In this region, lattice QCD provides us with a non-perturbative tool 
to calculate physical quantities of hadrons from first principles.
A lot of works have been devoted over two decades to numerical
simulations of lattice QCD, and gradually revealed that numerical
method is a powerful and practical tool.

Lattice QCD has been applied to calculations of the hadronic spectra
including glueballs and hybrids, the hadronic matrix elements of
operators of importance to weak decays and other hadronic
quantities such as the pion nucleon sigma term and the proton spin
carried by the quarks.
High temperature QCD has been an important research area of lattice QCD.
The topological structure of the QCD vacuum has also been investigated,
because it may provide a qualitative understanding of the mechanism
responsible for confinement and spontaneous breakdown of chiral symmetry.

Among various subjects pursued by lattice QCD, deriving the light hadron
spectrum is a key step, because it will be a fundamental confirmation
of QCD at low energies and give us confidence when we apply lattice
QCD to calculations of other physical quantities.
Considerable progress has been achieved in this field
over the past two years \cite{ref:review}.

\section{SYSTEMATIC ERROR IN LATTICE CALCULATIONS}
The first attempt to calculate the light hadron spectrum was made
in 1981 \cite{ref:HP,ref:W}.
Numerous studies devoted to this issue 
have revealed afterwards that we have to overcome several difficulties 
to obtain numerically precise results.

Because of the limitation of computer power, most of works to
calculate physical quantities of hadrons employ
{\it quenched approximation}, in which we ignore sea quark effects. 
One may expect that the quenching error is not so large, because 
the valence quark model describes qualitatively the observed spectrum.
However it it difficult to evaluate the magnitude of the error 
theoretically.
In order to justify the approximation, we first have to estimate 
its magnitude by comparing the quenched spectrum with experiment. 

In lattice QCD, the prediction for the real world can be obtained
only after we take the infinite volume and continuum limit.
Therefore we have to control systematic errors due to finite lattice
size and finite lattice spacing. 
In addition, because computational cost rapidly becomes very large 
as the quark mass decreases, one has to extrapolate results obtained
at heavy quark masses to the physical up and down quark masses.
Present typical calculations are still limited to the range of the quark mass 
$m_q \simgt$ 30 MeV.

The first step toward the goal of the spectrum calculation is therefore
to obtain the definitive result for the quenched QCD spectrum, by
controlling all systematic errors arising from extrapolations
mentioned above.

\section{QUENCHED QCD SPECTRUM}
\subsection{A history}
\begin{table}[t]
\newlength{\digitwidth} \settowidth{\digitwidth}{\rm 0}
\catcode`?=\active \def?{\kern\digitwidth}
\caption{Simulation parameters versus year.}
\label{tab:PrmVsYear}
\begin{tabular*}{\textwidth}{@{}l@{\extracolsep{\fill}}rrrrrr}
\hline
year &  1982   &  1985  & 1988  & 1993 & 1998 \\
\hline
$L_t$    &    16   &    24  &   50 & 50 & 56--112\\
$L_s$    &     8   & $\sim$ 12 & 16--24 & 24--36 & 32--64\\ 
$L_sa$ (fm)&   $\sim$ 0.8 & $\sim$ 1.2 & 1.4--1.7  & 2.0--2.5 & 3.0 \\
$a$ (fm) &  0.14--0.1 & 0.14--0.1 & 0.14--0.1 & 0.1--0.07 & 0.1--0.05 \\
$m_q$ (MeV) & $>120$ & $>120$  & $>40$ & $>40$ & 160--30 \\ 
$m_{\rm PS}/m_{\rm V}$ & $> 0.7$ & $> 0.7$ 
& $ > 0.5$ & $ $\simgt$ 0.5$ & 0.75--0.4 \\
statistics & 20 & 20 & 50 & 200 & 150--800 \\
\hline
\end{tabular*}
\end{table}
Instead of reviewing a long history of quenched spectrum calculations,
we reproduce in Table \ref{tab:PrmVsYear} typical simulation 
parameters versus year.
$L_t$ ($L_s$) in the table is the number of points in the temporal (spatial) 
direction and $a$ is the lattice spacing. 
Before 1993, physical lattice sizes were limited to the range $La \simlt$
2 fm, which was smaller than a typical size of hadrons, and therefore 
works made before 1993 were in an exploratory stage.
In the table, we reproduce, in addition to the quark mass,
corresponding value of the pseudo-scalar vector 
mass ratio ($m_{\rm PS}/m_{\rm V}$),
which is often quoted to show how close we are to the physical up and 
down quark masses, $m_q \approx$ 5 MeV.  
The ratios were limited to the range $m_{\rm PS}/m_{\rm V} \simgt 0.5$,
($m_q \simgt$ 40 MeV),  
which was far from $m_{\rm PS}/m_{\rm V} = 0.176$ at the physical point.

\begin{figure}[t]
\begin{minipage}[t]{77mm}
\centerline{\epsfxsize=77mm \epsfbox{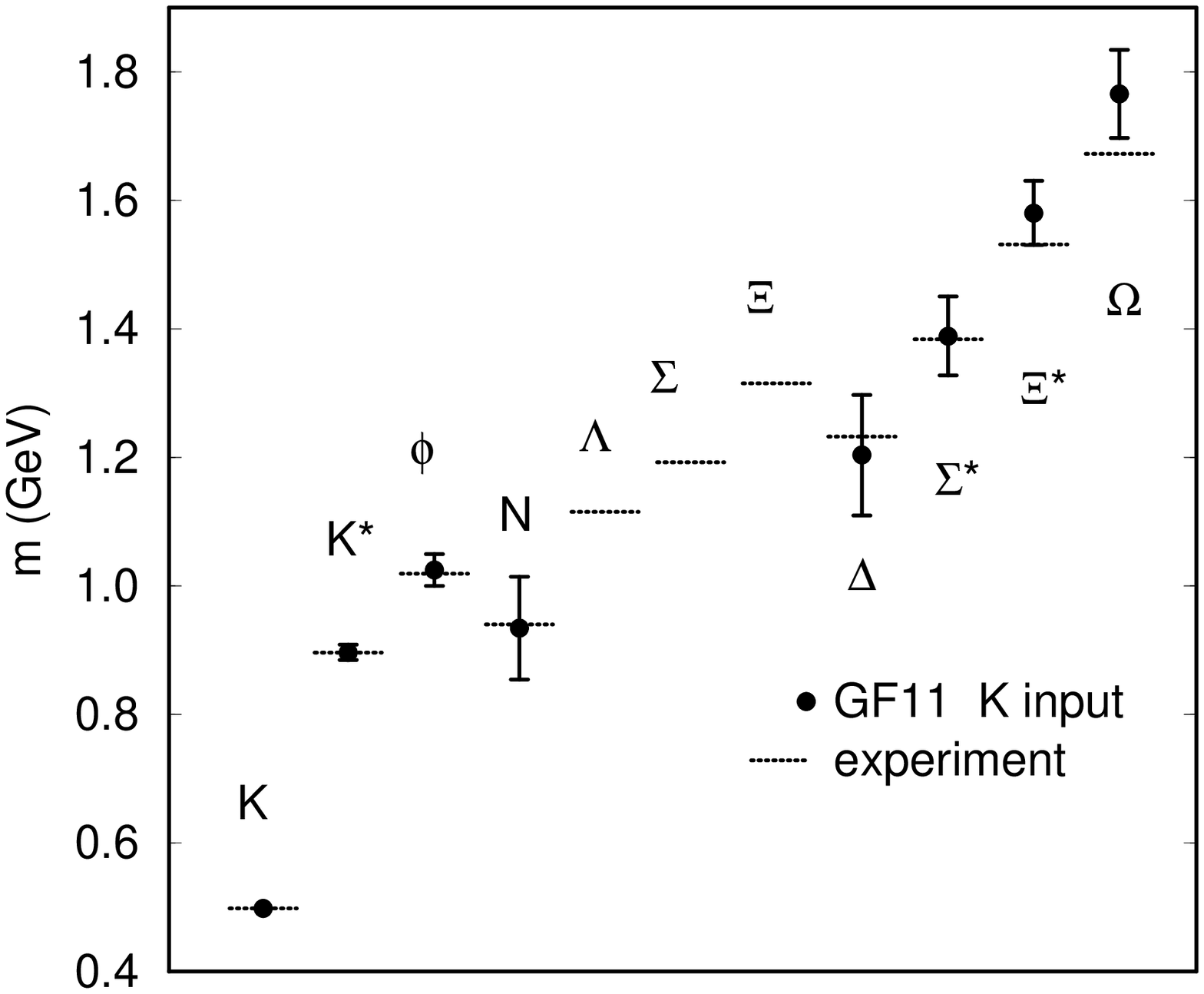}}
\vspace{-5mm}
\caption{Quenched QCD spectrum reported by the GF11 
collaboration \protect\cite{ref:GF11}.}
\label{fig:GF11}
\end{minipage}
\hspace{\fill}
\begin{minipage}[t]{77mm}
\centerline{\epsfxsize=77mm \epsfbox{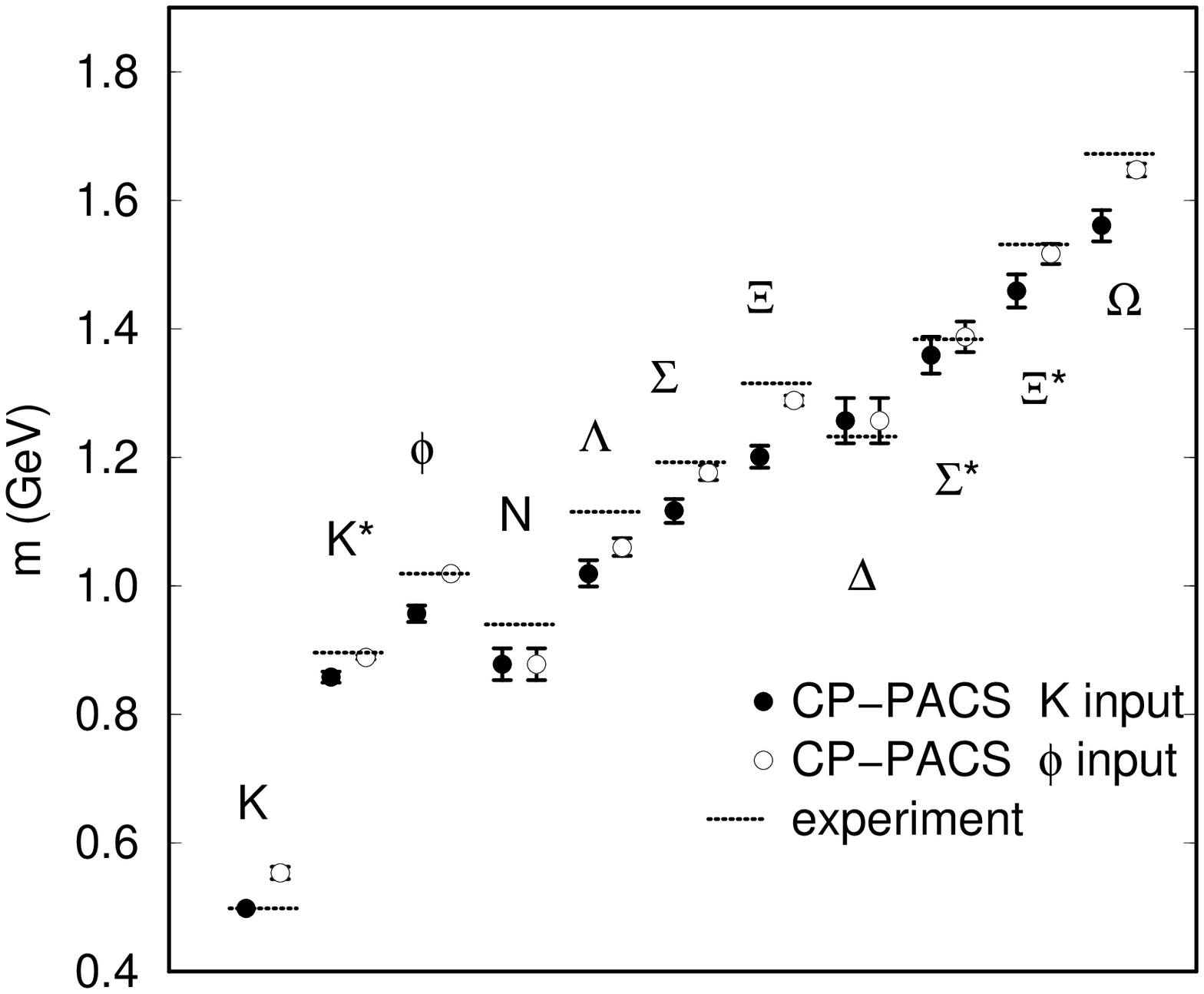}}
\vspace{-5mm}
\caption{The most recent result of the quenched QCD 
spectrum \protect\cite{ref:CPPACS-Q,ref:CPPACS-R}.}
\label{fig:CPPACSspec}
\end{minipage}
\end{figure}

A turning point of the calculation of the quenched spectrum was marked
by the GF11 collaboration around 1993 \cite{ref:GF11}.
They made simulations at three lattice spacings ranging over $a=$ 
0.1--0.07 fm on lattices with $La \approx$ 2.3 fm. 
After making continuum extrapolations and a finite size correction,
they obtained results shown in Fig.\ref{fig:GF11}.
In their calculation, masses of $\pi$, $\rho$ and $K$ were taken as input 
to fix the lattice spacing, the physical (and degenerate) up and down
quark masses and the strange quark mass. 
Although the entire spectrum was not covered, their results turned out
to be consistent with experiment, within one standard deviation,
or 1--9 \%.

In spite of this encouraging result, there was no further work to give
results for the entire spectrum with errors convincingly better than 5 \%.
In particular, the question of how the quenched QCD spectrum
deviates from experiment was not answered.

\subsection{State-of-the-art calculation}
A report from the CP-PACS collaboration in 1998 \cite{ref:CPPACS-Q,ref:CPPACS-R}
changed the situation.
They made an extensive simulation to
obtain the quenched spectrum with errors of 1--2 \% for mesons 
and 2--3 \% for baryons.
Their results have sufficient accuracy to discuss how the quenched spectrum
deviates from the experimental spectrum.
In Fig.\ref{fig:CPPACSspec} are summarized their spectrum result.

First, we notice that the discrepancy between the CP-PACS result 
and experiment is less than 11 percent. 
This means that the quenching error is of order 10 \%, 
which is not so large as we have expected.
This observation is important because we may expect that the quenching
error is generally of a similar order also for other physical quantities.

In Fig.\ref{fig:CPPACSspec}, one can see statistically significant
and systematic deviation of the quenched spectrum from experiment, 
amounting to 7$\sigma$ for some particles.
How the quenched spectrum deviates is summarized as follows.
If one uses $K$ meson mass as input to fix the strange quark mass, 
1) masses of vector mesons $m_{K^*}$ and $m_\phi$ are smaller than 
experiment, 2) octet baryon masses are systematically smaller than experiment
and 3) decuplet baryon mass splitting is smaller than experiment.
If one uses $m_\phi$ instead of $m_K$ as input, $m_{K^*}$
appears consistent with experiment and the discrepancies for baryon
masses are much reduced. However, $m_K$ turns out to be much higher.
In other words, the meson hyperfine splitting remains smaller than experiment.

In summary, the CP-PACS results show that
the strange quark mass cannot be tuned in quenched QCD
so that all strange hadron masses are in agreement with experiment.

\subsection{Possible systematic error}
Controlling systematic errors is a basic element in spectrum calculations.
We discuss in some detail the CP-PACS attempt to reduce them
so that the reader can understand what was improved and how large 
the remaining errors are.

\begin{figure}[t]
\begin{minipage}[t]{77mm}
\centerline{\epsfxsize=77mm \epsfbox{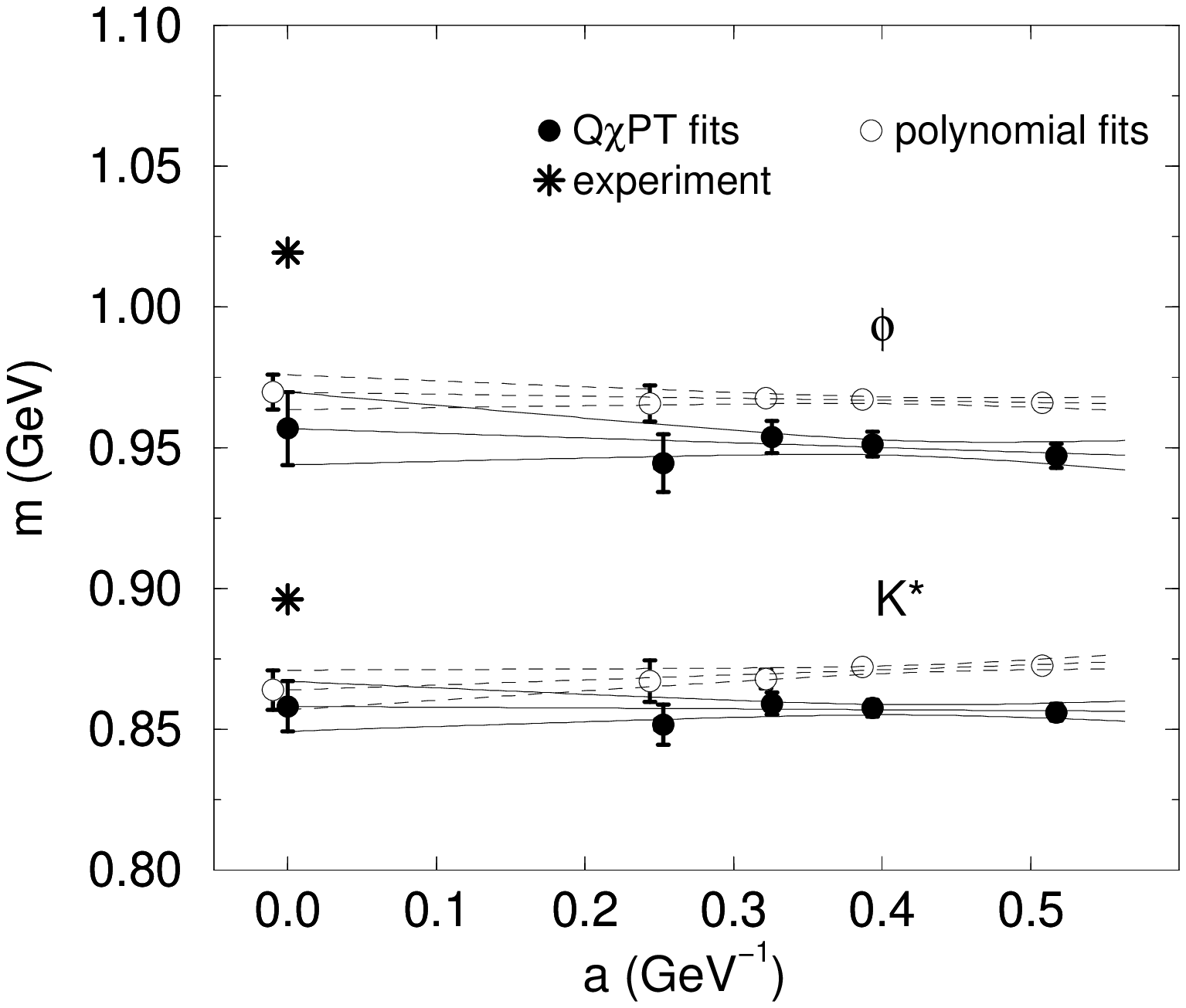}}
\vspace{-5mm}
\caption{Continuum extrapolations of meson masses with $m_K$ used as input.
Results from two different chiral extrapolations are reproduced.
See text for details.}
\label{fig:MesonCont}
\end{minipage}
\hspace{\fill}
\begin{minipage}[t]{77mm}
\centerline{\epsfxsize=77mm \epsfbox{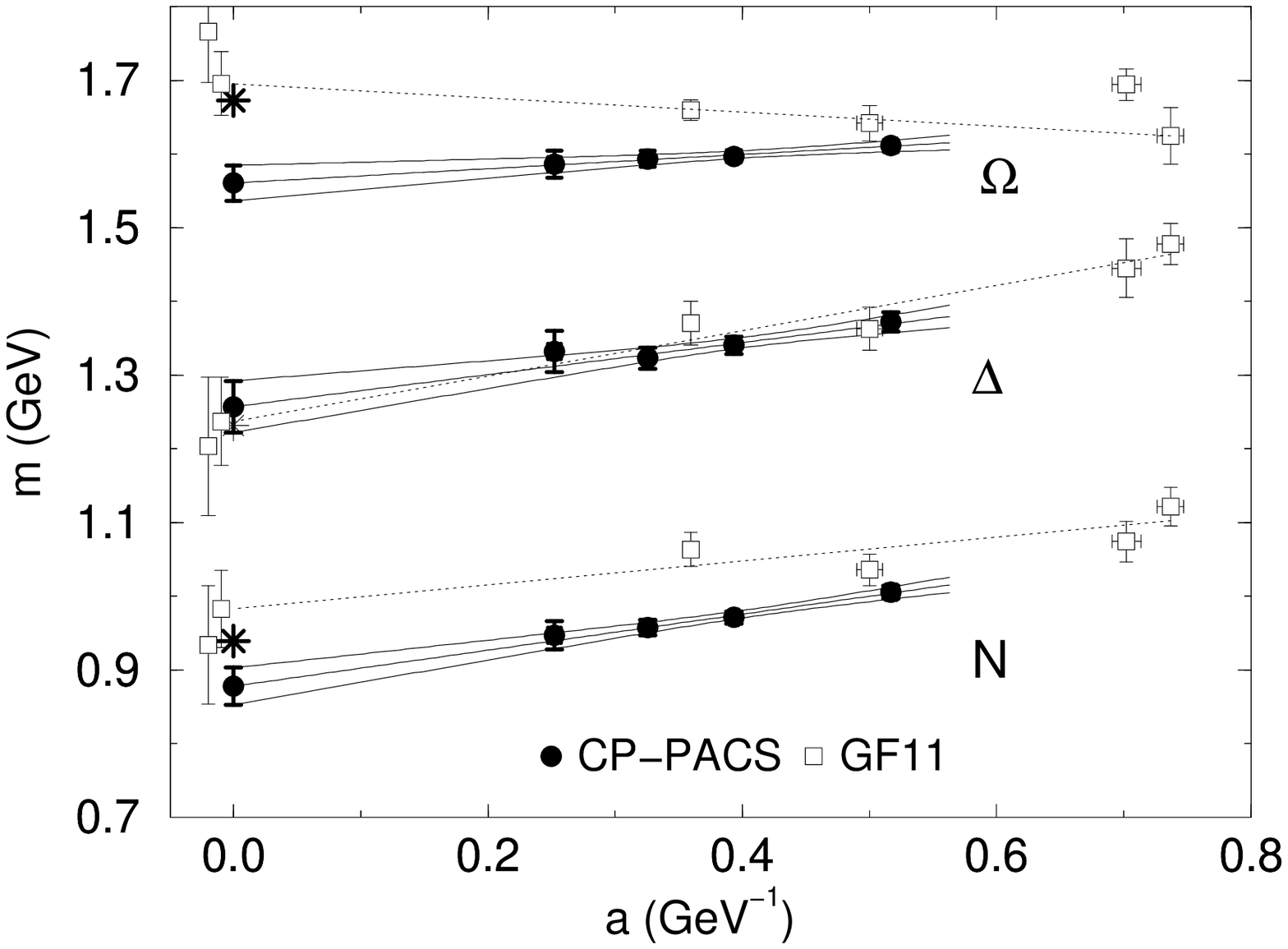}}
\vspace{-5mm}
\caption{Continuum extrapolations of baryon masses with $m_K$ used as input.
The CP-PACS results are compared with GF11's.} 
\label{fig:BaryonCont}
\end{minipage}
\end{figure}

They performed simulations on four lattices with $La \simgt$ 3 fm.
For these lattices, finite size effect is estimated to be smaller
than 0.4 \% for the worst case.

The CP-PACS collaboration employed the Wilson's formulation of quarks 
on a lattice. For continuum extrapolations with the Wilson action,
the leading correction to the continuum
value is proportional to the lattice spacing:
\begin{equation}
m(a) = m(a=0) \times [ 1 + C_1 a + (C_2 a)^2 + \cdots].
\end{equation}
CP-PACS data are fitted well by a linear function of 
the lattice spacing, as shown in Figs. 
\ref{fig:MesonCont} and \ref{fig:BaryonCont}.
The coefficient of the linear term $C_1$ ranges from 
0 to 0.28 GeV. One can evaluate the magnitude of higher order terms, 
assuming that $C_1 \approx C_2$.
We find that the quadratic term has an effect of at most 1 \%
at $a=0.075$ fm, the central value of the range of their lattice spacings.
Therefore finite lattice spacing effect which may not be removed by linear
extrapolations is estimated to be 1\% or less.

\begin{figure}[t]
\begin{minipage}[t]{77mm}
\centerline{\epsfxsize=77mm \epsfbox{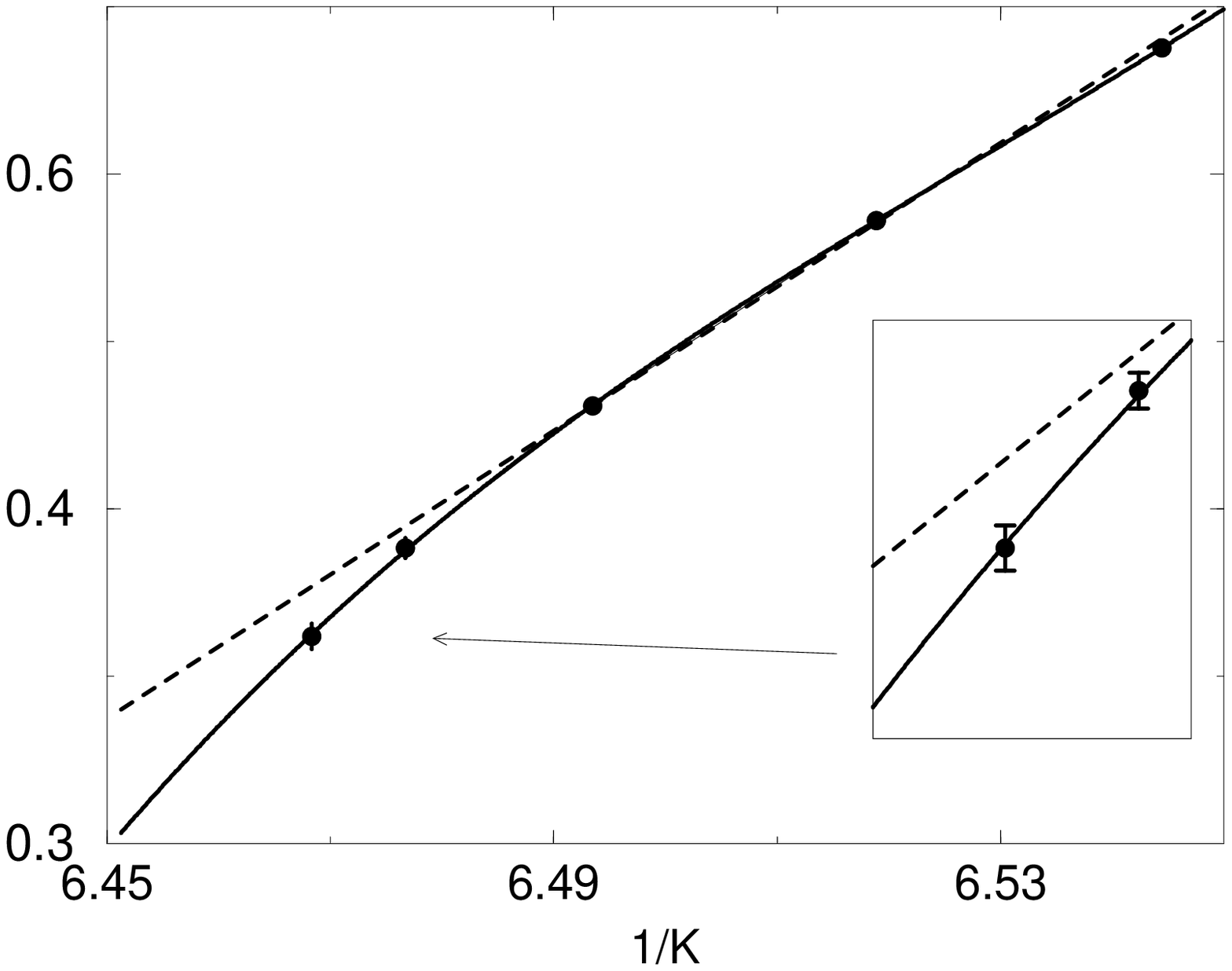}}
\vspace{-5mm}
\caption{Nucleon masses versus $1/K$. The quark mass $m_q$ is
proportional to $(1/K-1/K_c)/2$, where $K_c$ is a critical value.}
\label{fig:nucleon}
\end{minipage}
\hspace{\fill}
\begin{minipage}[t]{77mm}
\centerline{\epsfxsize=77mm \epsfbox{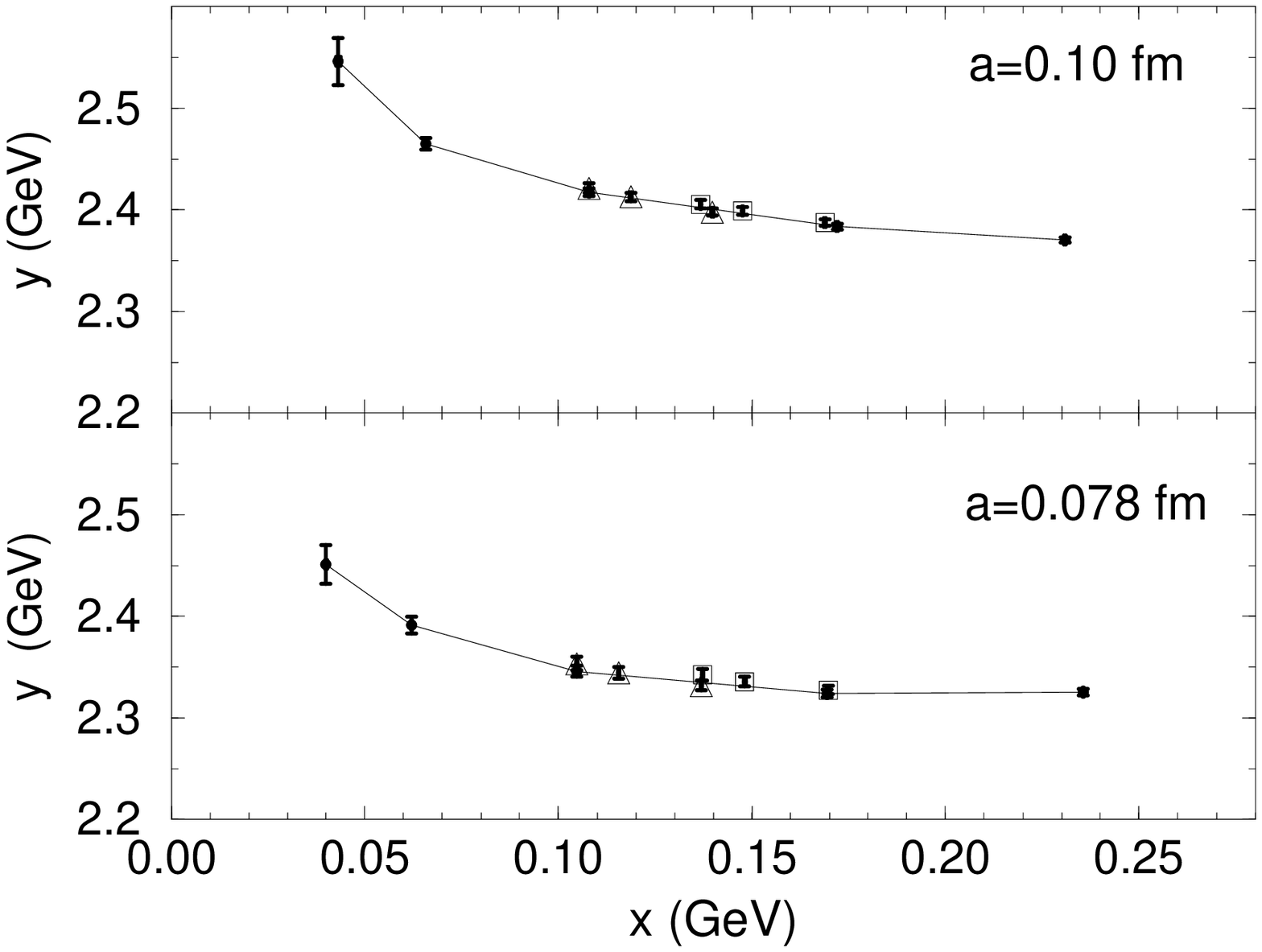}}
\vspace{-5mm}
\caption{$y=m_{\rm PS}^2/(m_{q_1}+m_{q_2})$ versus
$x=m_{q_1}+m_{q_2}$. Filled (open) symbols represent
degenerate (non-degenerate) data.}
\label{fig:qchpt}
\end{minipage}
\end{figure}

Chiral extrapolation is the most delicate issue in controlling
systematic errors. 
The CP-PACS collaboration performed simulations at five quark masses 
corresponding to $m_{\rm PS}/m_{\rm V}$ = 0.75, 0.7, 0.6, 0.5 and 0.4.
Previous studies with the Wilson quark action were limited to 
the range $m_{\rm PS}/m_{\rm V} \simgt$ 0.5, or $m_q \simgt$ 40 MeV.
Data at the smallest quark mass, $m_q \approx$ 30 MeV,
are indeed important for reliable chiral extrapolations.
For example, the nucleon mass as a function of the quark mass shows 
a clear negative curvature, as shown in Fig.\ref{fig:nucleon},
which may not be uncovered if the data at the smallest 
quark mass were lacking: 
Indeed, the GF11 collaboration made a linear chiral extrapolation.
Consequently, hadron masses at finite lattice spacing and hence
continuum extrapolations are quite different from those of GF11.
See Fig.\ref{fig:BaryonCont} for examples.

Theoretically, chiral extrapolations in quenched QCD are more
complicated than in full QCD.
The quenched chiral perturbation theory (Q$\chi$PT) \cite{ref:QCHPT}
predicts characteristic singularities in hadron masses in the chiral limit.
Therefore we first have to check numerically the validity of the
prediction. 
The CP-PACS collaboration extensively investigated this issue and found
several evidences for the existence of the quenched chiral singularities.
We reproduce one example. In full QCD, the relation 
$m_{\rm PS}^2 \propto m_q$ holds.  
In quenched QCD, however, $m_{\rm PS}^2$ has a logarithmic singular
term and the ratio $m_{\rm PS}^2/m_q$ diverges toward the chiral limit.
As Fig.\ref{fig:qchpt} shows, the CP-PACS data exhibits 
a clear increase of the ratio. 

The CP-PACS collaboration employed Q$\chi$PT mass 
formulae\cite{ref:QCHPT,ref:QCHPTM} for chiral extrapolations.
They also repeated the whole analysis employing conventional
polynomials in quark masses to investigate the effect of choosing
totally different chiral ansatz.
They found that the difference in the continuum limit was 1.5 \%
for the worst case, which was only a 1.5 $\sigma$ effect.
We reproduce their comparison of meson masses in Fig.\ref{fig:MesonCont}.

From these considerations, we conclude that 
the CP-PACS collaboration successfully reduced systematic errors 
to the magnitude of their statistical ones. 
Because the deviation from experiment 
amounts to 4--7 $\sigma$, taking account of systematic errors does
not change their conclusions in any significant ways.

Although the CP-PACS results are very convincing, a crosscheck is of
course necessary.
Two years ago, the MILC collaboration calculated \cite{ref:MILCKS}
the nucleon mass in the continuum limit using the Kogut-Susskind quark action,
another popular formulation of quarks on a lattice.
Their result $m_N = 964(35)$ MeV is slightly larger than experiment,
while $m_N = 878(25)$ MeV from the CP-PACS is smaller.
Understanding the difference, albeit only a 2.5 $\sigma$ effect, 
is important.
Also, the MILC collaboration calculated only the nucleon mass. 
It is desired to establish the entire spectrum for the
Kogut-Susskind action.

\section{IMPROVEMENT PROGRAM}
The CP-PACS result above was obtained by a full use of one of the fastest
computers in the world, CP-PACS, for about one year. 
Such a huge computational cost may make us pessimistic in doing realistic
calculations in full QCD.
Roughly speaking, computational cost for full QCD simulations is 
100--1000 times larger than that for quenched QCD.
Even in quenched QCD, there are many problems which need much more
computer time than the simple spectrum calculation.

\begin{figure}[t]
\begin{minipage}[t]{77mm}
\centerline{\epsfxsize=77mm \epsfbox{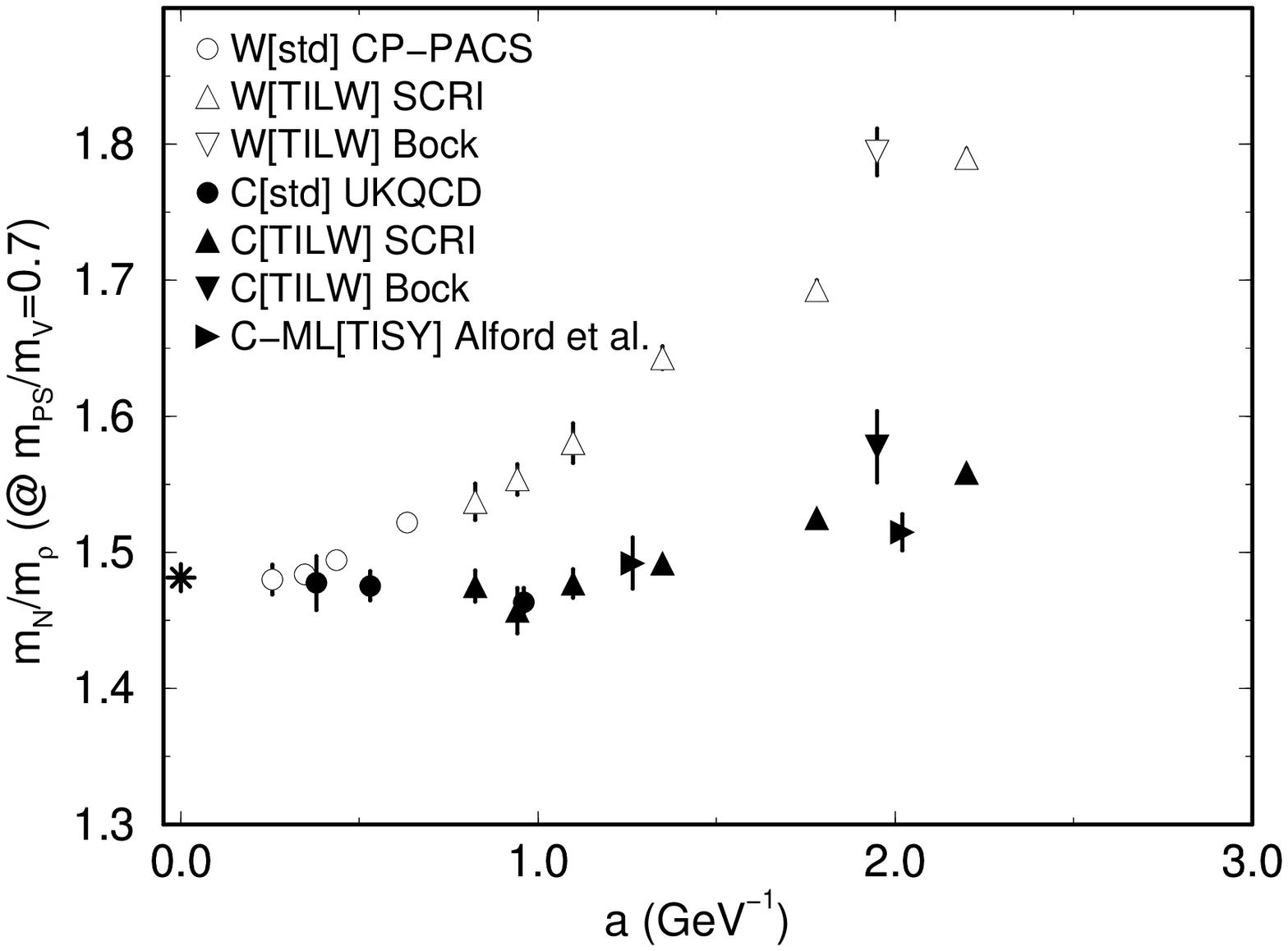}}
\vspace{-5mm}
\caption{Comparison of $m_N/m_\rho$ at $m_{\rm PS}/m_{\rm V}$ =0.7
for the Wilson (W) and clover (C) actions in quenched QCD.}
\label{fig:impQ}
\end{minipage}
\hspace{\fill}
\begin{minipage}[t]{77mm}
\centerline{\epsfxsize=77mm \epsfbox{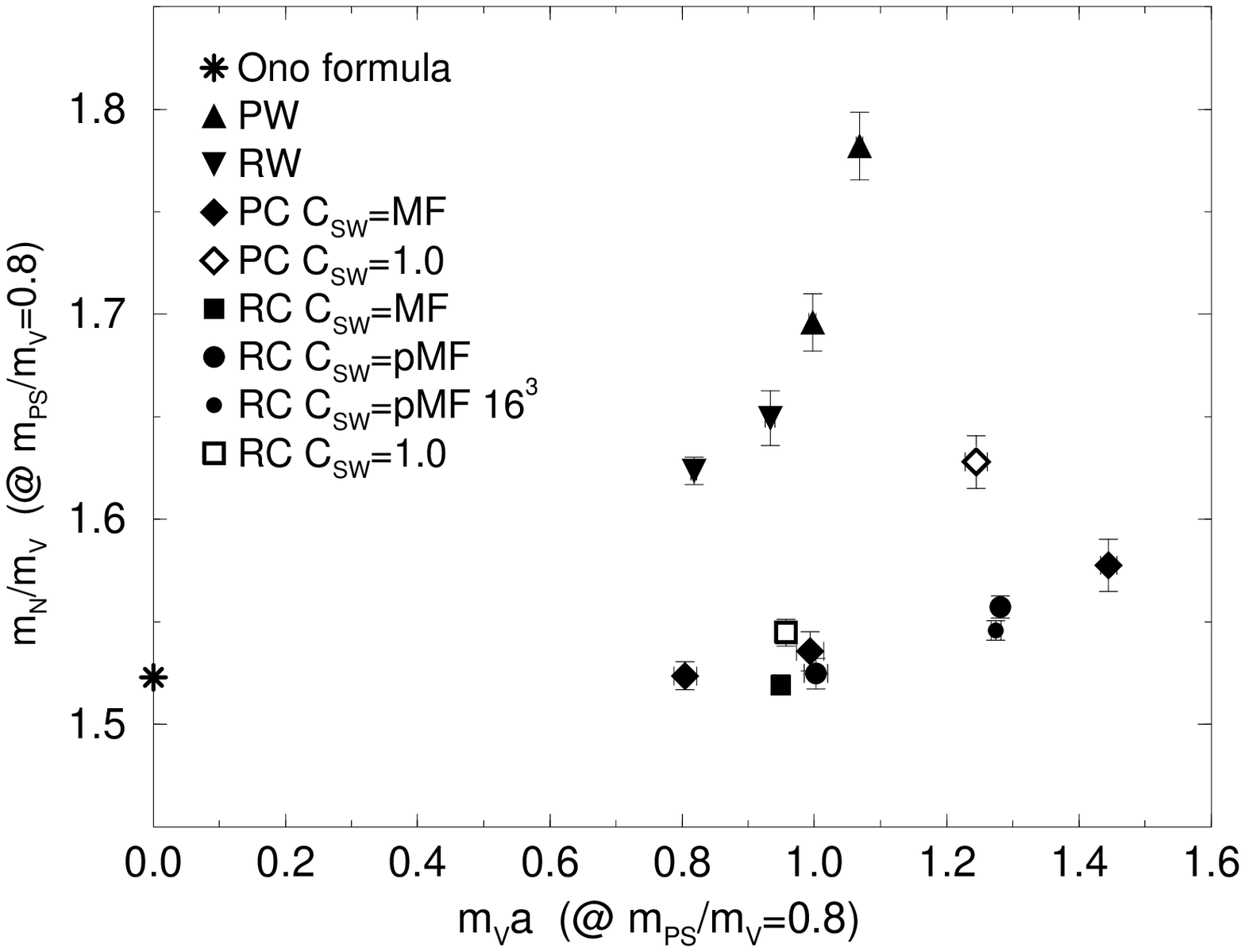}}
\vspace{-5mm}
\caption{$m_N/m_\rho$ in $N_f=2$ full QCD \protect\cite{ref:CPPACS-compara}.
Abbreviations: P:standard, R:improved for gluons, W:Wilson, C:Clover for quarks.} 
\label{fig:impF}
\end{minipage}
\end{figure}

A breakthrough to this situation may be given by improving lattice actions.
The basic idea is as follows.
Let us recall how we estimate the magnitude of the systematic error 
due to finite lattice spacing.
Because the leading correction is $O(a)$ for the Wilson action, 
and we fit data with a linear function, the remaining error is
of $O(a^2)$, which turned out to be about 1\% at $a \approx$ 0.075 fm.
Now suppose that we can invent an improved lattice action 
which has in principle no $O(a)$ correction.
If one uses the action and performs a fit using a quadratic function of 
the lattice spacing, the remaining error is reduced to $O(a^3)$.
In this case, we can obtain values in the continuum limit 
with the same accuracy from simulations at $a \approx$ 0.15 fm.
This value is larger by a factor two than for the Wilson action.
We perform simulations with the physical lattice size $La$ being 
kept fixed, while computational cost depends on $L$, but not on $La$.
Because computational cost becomes large faster than the number of lattice
points $L^4$, computational cost is reduced by at least a factor 16 for 
the improved action.

Years ago, Symanzik argued a general method to improve lattice action, 
order by order in lattice spacing, by adding higher dimensional
terms to the original one \cite{ref:Symanzik}.
Sheikholeslami and Wohlert then derived the $O(a)$ improved 
action\cite{ref:clover} for the Wilson quark action.
The form of the improved action is given by 
\begin{equation}
S_{\rm Clover} = S_{\rm Wison} + 
a \times c_{\rm sw}(g^2) \cdot {\bar q} \sigma_{\mu\nu}F_{\mu\nu} q, 
\end{equation}
where $g$ is the gauge coupling and $F_{\mu\nu}$ is the field strength.
Because the lattice representation of $F_{\mu\nu}$ looks like a clover with 
four leaves, the action is often called ``clover action''.
The coefficient $c_{\rm sw}$ can be calculated perturbatively
or non-perturbatively.

\begin{figure}[t]
\begin{minipage}[t]{77mm}
\centerline{\epsfxsize=77mm \epsfbox{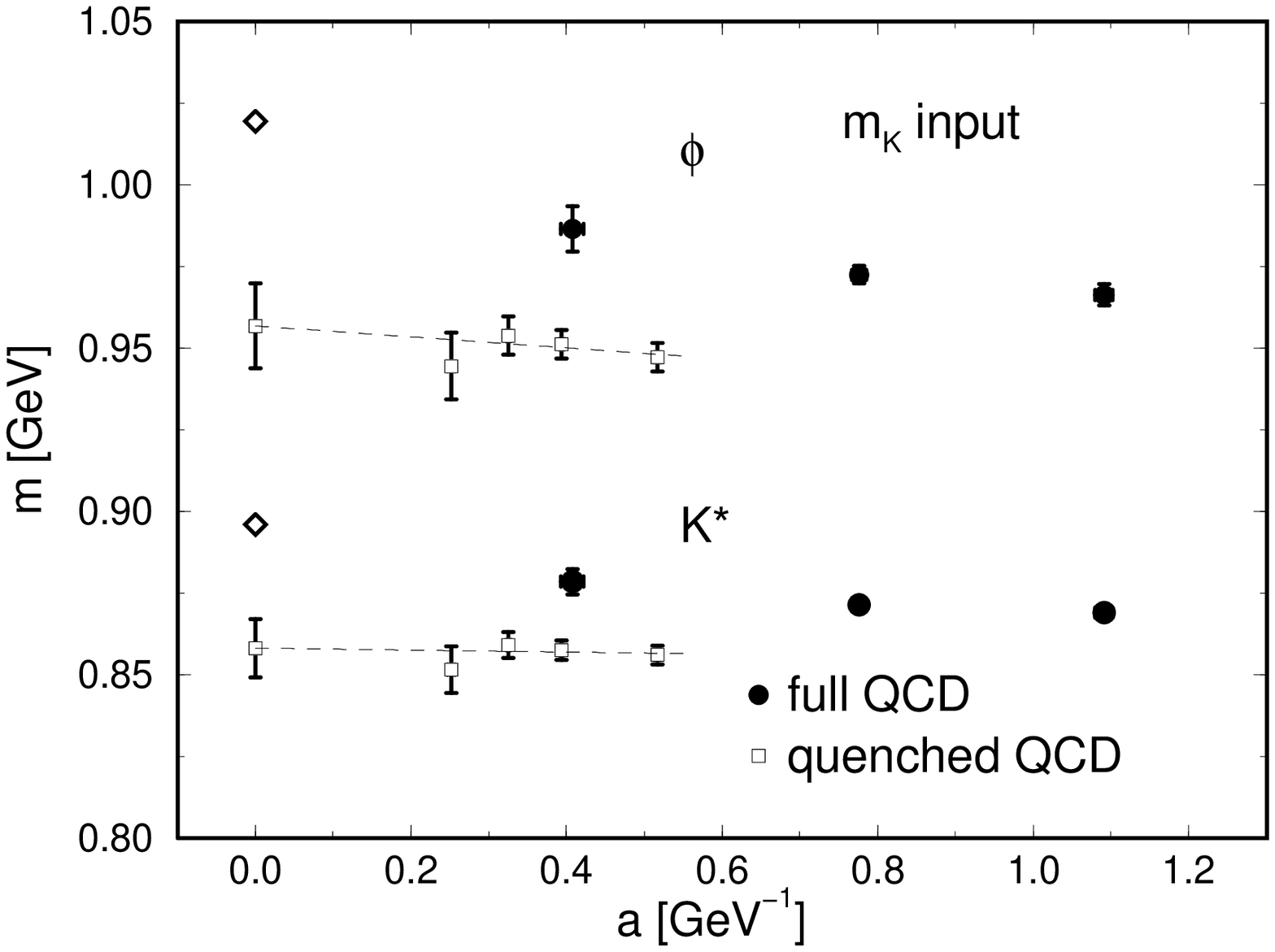}}
\vspace{-5mm}
\caption{Continuum extrapolations of vector meson masses in $N_f=2$
QCD, compared to those in quenched QCD.}
\label{fig:FMsnCont}
\end{minipage}
\hspace{\fill}
\begin{minipage}[t]{77mm}
\centerline{\epsfxsize=77mm \epsfbox{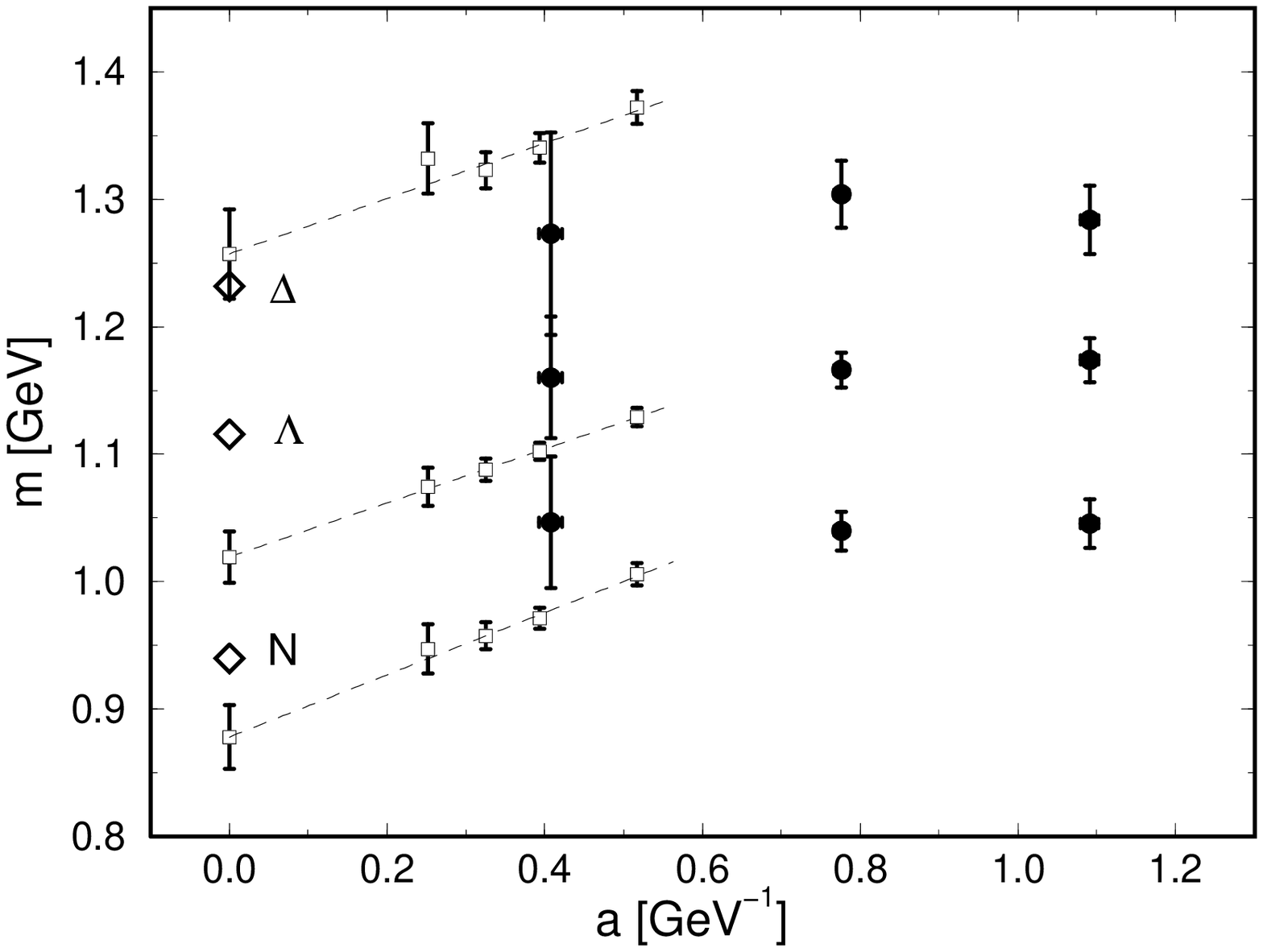}}
\vspace{-5mm}
\caption{The same as Fig.\protect\ref{fig:FMsnCont} for typical baryon masses.}
\label{fig:FBrnCont}
\end{minipage}
\end{figure}

Fig.\ref{fig:impQ} shows a compilation of the results for 
the nucleon to $\rho$ mass ratio ($m_N/m_\rho$) at $m_{\rm PS}/m_{\rm V} =$ 0.7
as a function of the lattice spacing.
We clearly observe that improving actions significantly reduces
scaling violation so that the ratio agrees with a phenomenological
value (the star in the figure) within 5\% already at $a\approx 0.4$ fm,
whereas the scaling violation for the Wilson action amounts to 20\% 
at the same lattice spacing.

Similar improvement is observed also 
for full QCD \cite{ref:CPPACS-compara}.
In Fig.\ref{fig:impF} are compared $m_N/m_\rho$ at $m_{\rm PS}/m_{\rm V} =$ 0.8
in two flavor QCD for four possible types of action combinations, 
the standard plaquette or an improved action for gluons, 
and the Wilson or the clover action for quarks.

\section{FULL QCD SPECTRUM}
The observation of improvement in full QCD encourages us to do
a systematic study of the full QCD spectrum using improved actions.
The CP-PACS collaboration is now working on this 
subject \cite{ref:CPPACS-R,ref:CPPACS-F}, employing
a renormalization group improved action for gluons and 
the clover action for quarks.
They simulate two degenerate sea quarks, 
identified with the up and down quarks, while the strange quark 
is treated in the quenched approximation.
Calculations are made at three lattice spacings in the range
$a \approx$ 0.2 -- 0.1 fm, keeping the physical lattice size at 
$La \approx$ 2.4 fm.
For chiral extrapolations and interpolations to the physical strange
quark mass, they use data at four sea quark masses in the range 
$m_{\rm PS}/m_{\rm V}$ = 0.8--0.6 and five valence quark masses 
with $m_{\rm PS}/m_{\rm V}$ = 0.8--0.5.

Our interest at this stage is how the discrepancy observed in quenched 
QCD is reduced, when one introduces two light dynamical quarks.
We reproduce in Figs.\ref{fig:FMsnCont} and \ref{fig:FBrnCont} 
the CP-PACS results as a function of the lattice spacing 
and compare them with the quenched spectrum.
We clearly observe that vector meson masses extrapolate to values
noticeably clover to experiment than those for quenched QCD.
The remaining discrepancy might be due to the quenching effect of 
the strange quark.
Sea quark effects in baryon sector are less clear.
We observe small scaling violation for baryons, which is an encouraging result.
However, the mass results lie 5--10\% larger than experiment.
The discrepancy might be due to possible finite size effects
($La\approx$ 2.4 fm for the full QCD data, while $La\approx$ 3.0 fm for
the quenched QCD data), or chiral extrapolation uncertainties
(the smallest quark mass for full QCD corresponds to $m_{\rm PS}/m_{\rm V}
\approx 0.5$, while $\approx 0.4$ for quenched QCD).
Further study is necessary to clarify sea quark effects in baryon masses.

\section{CONCLUSIONS}
In summary, we now have precise results for the quenched QCD spectrum
which reproduces the observed spectrum with discrepancies of order 10 \%.
Simulations with two dynamical quarks show indications that 
the discrepancies in quenched QCD for the strange hadron spectrum are reduced,
although we still have to treat the strange quark in the quenched approximation.

In my opinion, successful application of improved actions and 
fast development of computer power have combined to open a new era
of lattice QCD simulations.
A realistic calculation of full QCD will be achieved in the near
future, which would shed new light on hadron physics at low energies.

\vspace{2mm}
I thank members of CP-PACS group for their collaboration.
I am grateful to Y.Iwasaki and A.Ukawa for valuable suggestions 
on the manuscript.
This work is supported in part by the Grant-in-Aid No.09304029 
of Ministry of Education and University of Tsukuba Project Research.

\end{document}